\documentclass[floatfix, noshowpacs, preprintnumbers, twocolumn, amsmath, amssymb, aps, prb, superscriptaddress]{revtex4-2}

\pdfoutput=1

\usepackage{graphicx, xcolor}
\usepackage{soul}
\usepackage{mathrsfs}
\usepackage{float}
\usepackage{subfloat}

\usepackage{amsmath,amscd,amsbsy,amssymb,latexsym,url,bm,amsthm}
\usepackage[vlined,boxed,commentsnumbered,linesnumbered,ruled]{algorithm2e}

\newcommand{\RNum}[1]{\uppercase\expandafter{\romannumeral #1\relax}}

\renewcommand{\i}{\mathrm{i}}

\begin{document}
\title{Codesigned counterdiabatic quantum optimization on a photonic quantum processor}

\author{Xiao-Wen Shang}
\altaffiliation{These authors contributed equally to this work.}
\affiliation{Center for Integrated Quantum Information Technologies (IQIT), School of Physics and Astronomy and State Key Laboratory of Advanced Optical Communication Systems and Networks, Shanghai Jiao Tong University, Shanghai 200240, China}
\affiliation{TuringQ Co., Ltd., Shanghai 200240, China}

\author{Xuan Chen}
\altaffiliation{These authors contributed equally to this work.}
\affiliation{Center for Integrated Quantum Information Technologies (IQIT), School of Physics and Astronomy and State Key Laboratory of Advanced Optical Communication Systems and Networks, Shanghai Jiao Tong University, Shanghai 200240, China}

\author{Narendra N. Hegade}
\affiliation{Kipu Quantum GmbH, Greifswalderstrasse 212, 10405 Berlin, Germany}

\author{Ze-Feng Lan}
\affiliation{Center for Integrated Quantum Information Technologies (IQIT), School of Physics and Astronomy and State Key Laboratory of Advanced Optical Communication Systems and Networks, Shanghai Jiao Tong University, Shanghai 200240, China}

\author{Xuan-Kun Li}
\affiliation{Center for Integrated Quantum Information Technologies (IQIT), School of Physics and Astronomy and State Key Laboratory of Advanced Optical Communication Systems and Networks, Shanghai Jiao Tong University, Shanghai 200240, China}

\author{Hao Tang}
\email{htang2015@sjtu.edu.cn}
\affiliation{Center for Integrated Quantum Information Technologies (IQIT), School of Physics and Astronomy and State Key Laboratory of Advanced Optical Communication Systems and Networks, Shanghai Jiao Tong University, Shanghai 200240, China}\

\author{Yu-Quan Peng}
\affiliation{TuringQ Co., Ltd., Shanghai 200240, China}

\author{Enrique Solano}
\affiliation{Kipu Quantum GmbH, Greifswalderstrasse 212, 10405 Berlin, Germany}

\author{Xian-Min Jin}
\email{xianmin.jin@sjtu.edu.cn}
\affiliation{Center for Integrated Quantum Information Technologies (IQIT), School of Physics and Astronomy and State Key Laboratory of Advanced Optical Communication Systems and Networks, Shanghai Jiao Tong University, Shanghai 200240, China}
\affiliation{TuringQ Co., Ltd., Shanghai 200240, China}
\affiliation{Chip Hub for Integrated Photonics Xplore (CHIPX), Shanghai Jiao Tong University, Wuxi 214000, China}

\maketitle

\textbf{Codesign, an integral part of computer architecture referring to the information interaction in hardware-software stack, is able to boost the algorithm mapping and execution in the computer hardware. This well applies to the noisy intermediate-scale quantum era, where quantum algorithms and quantum processors both need to be shaped to allow for advantages in experimental implementations. The state-of-the-art quantum adiabatic optimization algorithm faces challenges for scaling up, where the deteriorating optimization performance is not necessarily alleviated by increasing the circuit depth given the noise in the hardware. The counterdiabatic term can be introduced to accelerate the convergence, but decomposing the unitary operator corresponding to the counterdiabatic terms into one and two-qubit gates may add additional burden to the digital circuit depth. In this work, we focus on the counterdiabatic protocol with a codesigned approach to implement this algorithm on a photonic quantum processor. The tunable Mach-Zehnder interferometer mesh provides rich programmable parameters for local and global manipulation, making it able to perform arbitrary unitary evolutions. Accordingly, we directly implement the unitary operation associated to the counterdiabatic quantum optimization on our processor without prior digitization. Furthermore, we develop and implement an optimized counterdiabatic method by tackling the higher-order many-body interaction terms. Moreover, we benchmark the performance in the case of factorization, by comparing the final success probability and the convergence speed. In conclusion, we experimentally demonstrate the advantages of a codesigned mapping of counterdiabatic quantum dynamics for quantum computing on photonic platforms.}
\section*{Introduction}

Over the past decade, quantum computers have been developed on various platforms, $e.g.$,  superconducting qubits \cite{barends2016digitized,wendin2017quantum,xbxuchip2022}, trapped ions \cite{leibfried2003quantum,blatt2012quantum}, and linear optics \cite{carolan2015universal,tillmann2013experimental}, among others. However, the quality of qubits and the fidelity of quantum gates are still inadequate to implement large and accurate quantum algorithms~\cite{daley2022practical,huang2022quantum,wu2021strong,zhong2020quantum,Schuld2015,grover1996fast,shor1994algorithms}, which may only be done with future fault-tolerant processors \cite{fowler2012surface,konno2024logical,michael2016new,cai2023quantum}. To accommodate the limited performance of current quantum devices, compromises are made to take advantage of quantum algorithms in the noisy intermediate-scale quantum (NISQ) era \cite{Gebhart2023,bharti2022noisy,hartmann2016quantum}. One of the widely adopted solutions is the variational quantum algorithm (VQA) which combines quantum evolution and classical optimization together \cite{cerezo2021variational,benedetti2021hardware,lubasch2020variational}. An appropriate ansatz is selected to construct a parameterized quantum circuit model, and classical optimizations are employed to minimize the cost function and search for the target state, which is typically the ground state. 

Specifically, the quantum approximate optimization algorithm (QAOA) is an efficient instance of the VQA and has been first implemented to solve combinatorial optimization problems \cite{farhi2014quantum}. Further researches extend the range of problems feasible for QAOA to solve \cite{vikstaal2020applying,hodson2019portfolio,brady2021optimal} and improve the performance of this algorithm \cite{zhu2022adaptive,bravyi2020obstacles,egger2021warm}. 
However, as the scale of the problems increases, QAOA becomes severely restricted by the growing circuit depth and the increasing number of parameters to be optimized. 

Another popular paradigm of quantum computing is adiabatic quantum computing, whose Hamiltonian shares a similar form with QAOA \cite{farhi2000quantum}. The problems AQC is facing, $e.g.$, requirement of long coherence time and slow evolution speed to ensure adiabaticity, are considerably reduced by introducing counterdiabatic (CD) driving, which suppresses the non-adiabatic process by adding a term related to the rate of change of the parameters and allows the system to evolve faster \cite{demirplak2003adiabatic, berry2009transitionless, del2013shortcuts}.
Similar techniques can also be applied to improve QAOAs, leading to a method that can be called counterdiabatic quantum optimization \cite{hegade2022digitized, chandarana2022digitized}.
However, implementing the exact CD protocol in many-body systems is computationally expensive and experimentally challenging due to the complexity of non-local terms that scale with system size. To address this challenge, several approximate protocols have been proposed, including the nested commutator method, which provides a practical way to approximate the CD terms~\cite{sels2017minimizing, claeys2019floquet}. This approach simplifies implementation while maintaining high fidelity, making it more suitable for near-term quantum devices, though it has yet to be experimentally verified.

\begin{figure}[t!]
\includegraphics[width=0.48\textwidth]{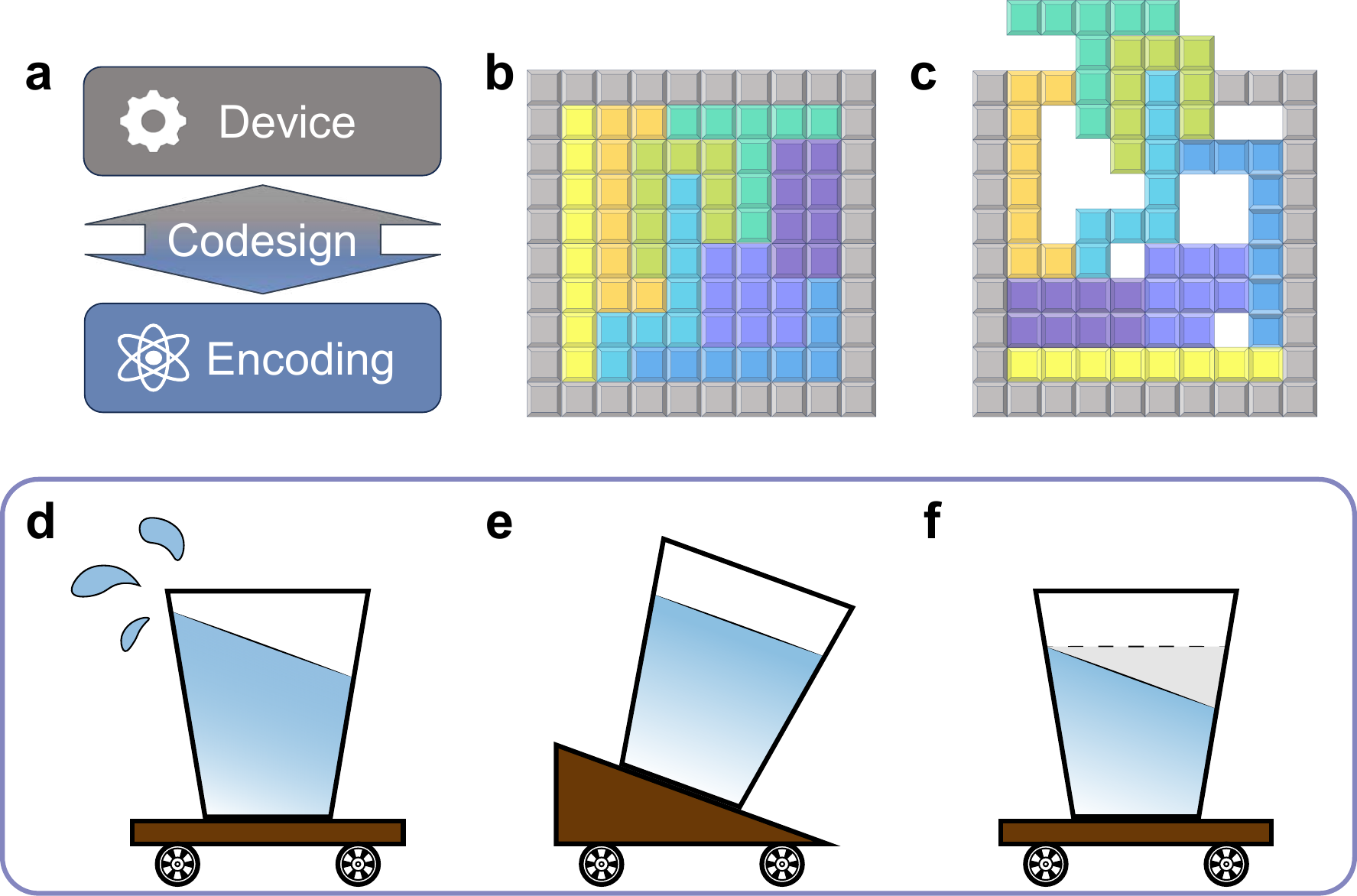}
\caption{\textbf{Illustration of codesign and the model of algorithms.} \textbf{a-c} Illustration of the codesign for quantum computing. \textbf{a} Codesign enables the optimal fulfillment of quantum computing tasks, by choosing proper encoding strategies that are compatible with the quantum devices so as to have the maximal utilization of the quantum resources. 
\textbf{b} By aligning the quantum resources with the problem requirements (building blocks in colors) with a proper encoding method, they are fully leveraged within the available quantum device (grey frame). \textbf{c} Without codesign, an improper encoding strategy fails to allocate all the required quantum resources within the available device. \textbf{d-f} We illustrate the quantum optimization process by transporting a glass of water. \textbf{d} A QAOA transfer has to move slowly, since moving quickly would have spilled water out of the glass. \textbf{e} A traditional counterdiabatic transfer is able to tilt the water glass to prevent spills during rapid movement. \textbf{f} An enhanced counterdiabatic transfer do not have to tilt the tray, since the water is low, implying the circuit ansatz is properly trimmed to be simplified. }
\label{fig:Fig1}
\end{figure}

From an experimental view, VQA has been firstly realized on a photonic quantum processor as an eigensolver \cite{peruzzo2014variational}. Successive works were also accomplished on superconducting platforms \cite{chen2020demonstration}, trapped ions \cite{hempel2018quantum}, and Rydberg atoms~\cite{ebadi2022quantum}. Recently, integrated photonic chips broadened the field of VQAs towards process tomography \cite{xue2022variational}, quantum unsampling~\cite{carolan2020variational}, and parameter estimation \cite{cimini2024variational}. The reconfigurability and long coherence time make photonic platform a potent candidate for applying VQAs \cite{dnliuchip2022,xdwangchip2022,Somhorst2203,li2024high}. However, some efficient algorithms are more complicated and demanding of experimental capabilities, leading to a lack of experimental demonstrations. Recently, a novel technique of hardware-software codesign has been proposed, allowing for the tailoring of abstract quantum circuits into customized versions that are highly effective, fully leveraging the strengths of the experimental platform~\cite{tomesh2021quantum}. Notably, the photonic platform's ability to perform global manipulation and execute unitary evolutions, while keeping precision and speed, makes it an ideal choice for implementing codesigned quantum optimization. This makes it possible to operate the counteradiabatic driving on a photonic circuit, but the corresponding codesigned algorithms are still to be put forward. 

Here, we propose a codesigned counterdiabatic quantum optimization (CCQO) algorithm, mapped onto a fabricated eight-mode photonic chip consisting of tunable Mach-Zehnder interferometers (MZIs). The MZI mesh, as a system-level processor, is explored to perform arbitrary unitary operators and optimization through global encoding. Unlike the traditional dual-rail encoding, global encoding codesigns the algorithm to fit our chip. This is done via direct matching of the evolution operators governed by the problem and CD Hamiltonians within our proposed CCQO. Furthermore, leveraging the nested commutator method, we enhance the performance of CCQO by extracting the coefficients of two-body operators from the problem Hamiltonian, while multiplying them by selected operators from a pool of local CD terms. This algorithm, denoted as CCQO-E by us, is experimentally demonstrated to solve a factorization problem with a high success probability even with a shallow circuit, outperforming CCQO and a codesigned QAOA (CQAOA). The results present a promising codesign strategy for multi-port quantum networks and quantum optimization algorithms, highlighting photonic chips as a powerful benchmark platform for quantum computation.

\section*{Results}

\subsection{Codesign of counterdiabatic quantum optimization alogorithm and photonic chip }
Codesign is an important concept in a wide range of research areas, with a core of collaborating between the software and the hardware. Recently, this strategy is adapted to designing quantum algorithms on the quantum devices by align the quantum device and encoding method to meet with the requirement of problems~\cite{tomesh2021quantum}, as shown in Fig.~\ref{fig:Fig1}a. The NISQ devices can only solve problems most efficiently when their quantum resources are well utilized by proper encoding, as illustrated in Fig.~\ref{fig:Fig1}b and Fig.~\ref{fig:Fig1}c, where the building blocks in colors and the grey frame are the quantum resources required for quantum algorithms and the current NISQ device that can be available, respectively. If without well codesign, the quantum resources required by the algorithm cannot be satisfied within the existing NISQ device box (see Fig.~\ref{fig:Fig1}c), which would then require a larger frame (more quantum devices) to cope with the algorithm. Specifically, in this work, we present the method of \emph{global encoding} on the photonic chip for the codesign of counterdiabatic driving quantum optimization algorithm. 

\begin{figure*}[t!]
\includegraphics[width=0.99\textwidth]{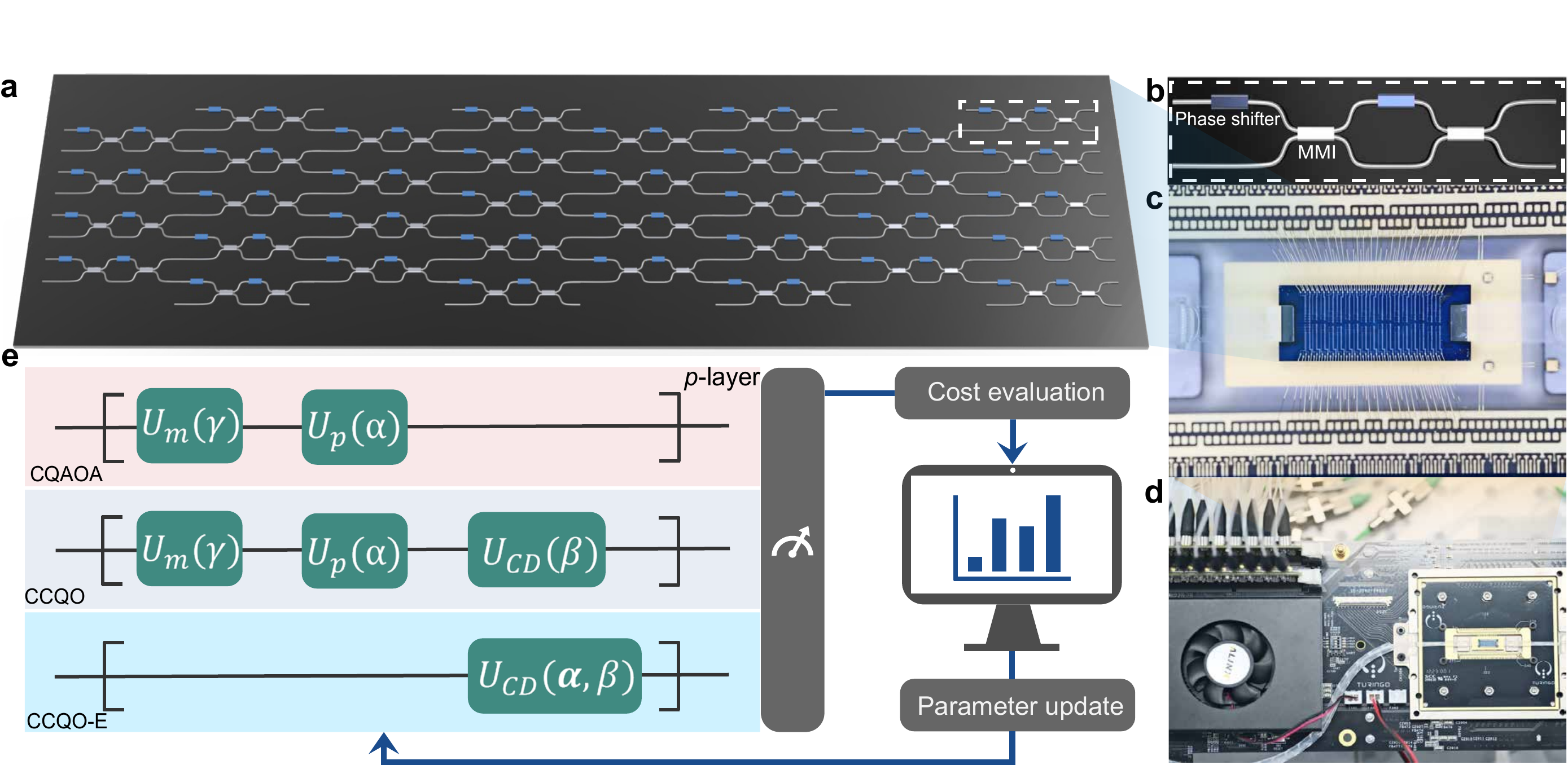}
\caption{\textbf{Experimental setup.} \textbf{a} Schematic of 8-mode photonic integrated chip (PIC) composed of a MZI mesh to realize arbitrary unitary evolution.  A single MZI on chip is framed with white dash line and zoomed in to show more details, as depicted in {\bf b}. Phase shifters (PS) are used to offer tunable phase difference between the two arms, and the multimode interfereometers (MMI) are designed to perform as a 50:50 beam splitter. \textbf{c} Detailed structure of the PIC. \textbf{d} Macroscopic picture of the development board packaging the PIC. \textbf{e} The on-chip optimization procedure of the codesigned algorithms. Depending on the algorithm, different terms of parameterized unitaries are adopted and loaded to the PIC. The output state is measured in the computational basis and used to evaluate the cost function. A personal computer (PC) is used to calculate the gradient and update the parameters. This optimization procedure is repeated until the cost function is minimized.
}
\label{fig:Fig2}
\end{figure*}

Due to the limitations of QAOA mentioned above, we consider an approximate CD protocol with the modification of nested commutators to get a high fidelity while maintaining the simplified implementations, which is more suitable for near-term quantum devices. The main idea behind counterdiabatic quantum optimization is to introduce a local counterdiatic term to the QAOA ansatz for the fast convergence to the ground state. An example of transferring a glass of water is sketched in Fig.~\ref{fig:Fig1}d-f. As illustrated by the tilted tray (Fig.~\ref{fig:Fig1}e),  
the additional CD Hamiltonian $H_{\mathrm{CD}}$ provides a compatible cushion to the problem Hamiltonian to allow for faster dynamics than QAOA. $H_{\mathrm{CD}}$ can be randomly chosen from an operator pool related to the problem Hamiltonian and is having similar interaction strengths as that of the problem Hamiltonian (see Methods). Between the problem Hamiltonian and the CD Hamiltonian, there is a mixing term $H_\mathrm{m}$ (see Methods). The parameterized unitary operator corresponding to the $p$-layered CCQO is given by
\begin{align}\label{eq4}
    U(\alpha, \beta, \gamma) = &U_{\mathrm{m}} (\beta_p )U_{\mathrm{CD}} (\alpha_p )U_{\mathrm{p}} (\gamma_p )U_{\mathrm{m}} (\beta_{p-1} )... \nonumber \\
    &U_{\mathrm{CD}} (\alpha_{p-1} )U_{\mathrm{p}} (\gamma_{p-1} )U_{\mathrm{m}} (\beta_1)U_{\mathrm{CD}} (\alpha_1)U_{\mathrm{p}} (\gamma_1),\notag \\
\end{align}
where the evolution operators are set as: $U_{\mathrm{m}}(\beta_p) = \exp (-\i\beta_p H_{\mathrm{m}})$, $U_{\mathrm{p}}(\gamma_p) = \exp (-\i\gamma_p H_{\mathrm{p}})$ and $U_{\mathrm{CD}} (\alpha_p)$ $= \exp (-\i\alpha_p H_{\mathrm{CD}})$.

However, this version of CD driving is still a challenge for many-body systems, since a many-body system often suffers from the short coherent time and the long-range control technique. As illustrated in Fig.~\ref{fig:Fig1}, further improvements can be made by dropping the many-body terms in the ansatz construction to reduce the requirement of fine tuning, $i.e.$ to avoid the need of tilting the tray. 
To enhance the counterdiabatic quantum optimization, instead of following the alternating operator sequence as in Eq. (\ref{eq4}), we construct the ansatz by randomly choosing the terms from the operator pool. For this particular instance the chosen ansatz is given by
\begin{equation}
\label{eq6}
	U_{\mathrm{CD}}(\alpha,\beta)=\exp\left[-\i\sum_i^N\alpha_ih_i^z\sigma_i^y\right]\exp\left[-\i\beta\sum_{i,j}J_{ij}\sigma_i^z\sigma_j^y\right],
\end{equation}
where $\alpha_i$, $\beta$ are the variational parameters, and $h_i^z$, $J_{ij}$ are the coefficients of one- and two-body interaction terms in the problem Hamiltonian. 

\begin{figure*}
\includegraphics[width=0.87\textwidth]{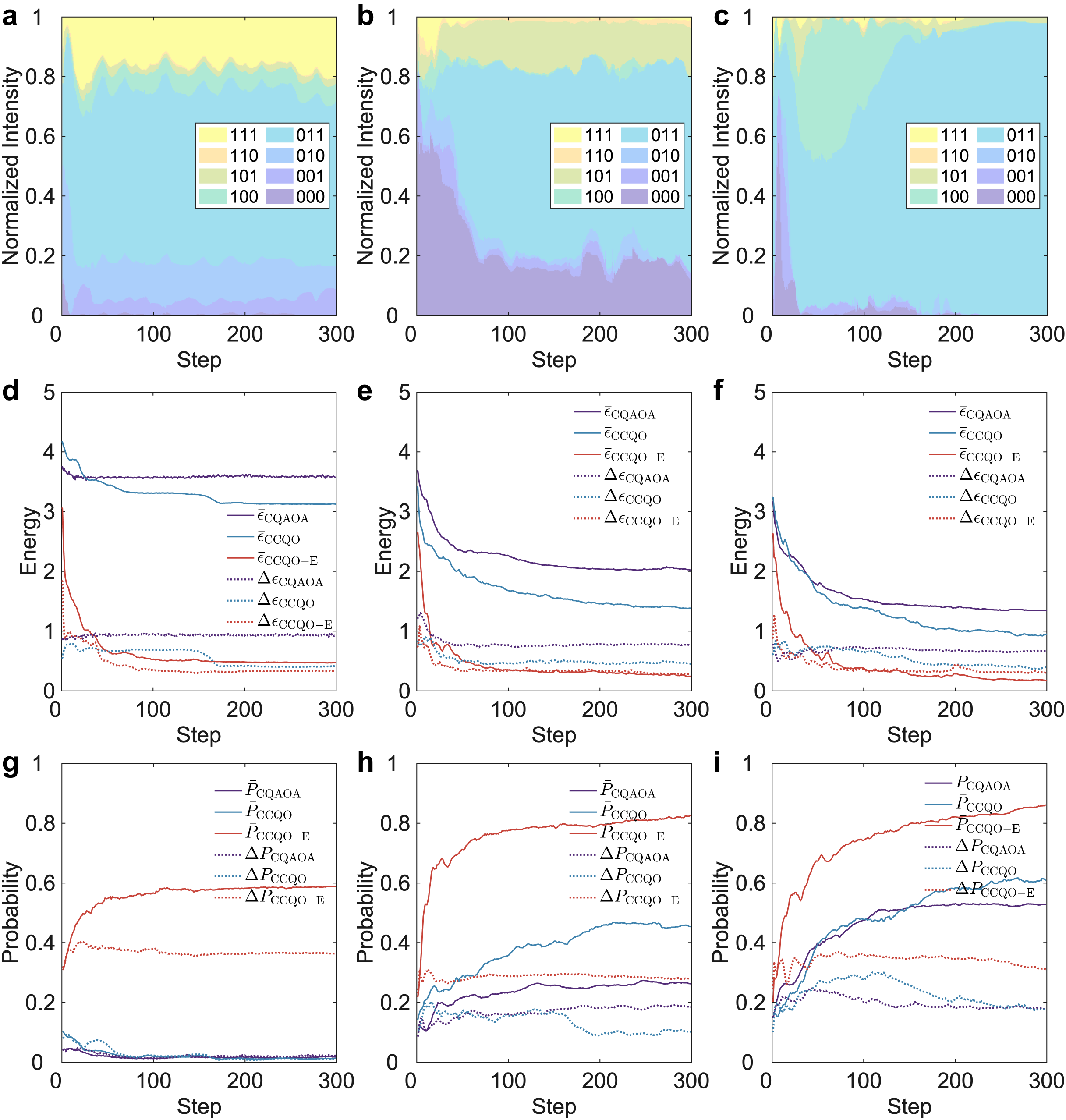}
\caption{\textbf{Experimental results of factoring the number 2893 using the three algorithms with different layers.} {\bf a, b} and {\bf c} Experimentally measured intensity of being in each of the eight states using CQAOA, CCQO and CCQO-E with 3 layers, respectively. The intensity is normalized to obtain the probability of being in each of the eight states. For each color, the vertical range of the color block indicates the probability of the corresponding state. The eight color blocks stacked together always gives a probability sum of 1. {\bf d, e} and {\bf f} Experimentally measured energy evolution during iterations with 1 layer, 2 layers, and 3 layers, respectively. {\bf g, h} and {\bf i} The corresponding experimentally measured success probability of obtaining the ground state $|011\rangle$ during iterations with 1 layer, 2 layers, and 3 layers, respectively. In {\bf d}-{\bf i}, the purple, blue and red lines are respectively the average results of 10 experiments using CQAOA, CCQO and CCQO-E. The dotted lines are the corresponding standard deviations. In each of the 10 experiments, the ansatz is initialized randomly with all parameters chosen in the range of $[-\pi,\pi)$.}
\label{fig:Fig3}
\end{figure*}

We map the operators in Eq.~(\ref{eq4}) and Eq.~(\ref{eq6}) in the photonic circuit composed of MZIs in a Clements configuration, as illustrated in Fig.~\ref{fig:Fig2}. 
The features of Eq.~(\ref{eq6}) is that the evolution is not governed by Hamiltonians for the term $\exp\left[-\i\sum_i^N\alpha_ih_i^z\sigma_i^y\right]$. Thus, the task of codesign becomes efficient mapping matrices on chip. To address this task, we propose the technique of \emph{global encoding}. As shown in Fig.~\ref{fig:Fig2}a, from top to bottom, each of the eight ports respectively encodes the state $|000\rangle$, $|001\rangle$, $|010\rangle$, $|011\rangle$, $|100\rangle$, $|101\rangle$, $|110\rangle$, and $|111\rangle$. The port where the photon is detected reveals the quantum state that the system is in. The on-chip control of the three-qubit state is based on the single MZI shown in Fig.~\ref{fig:Fig2}b, whose control system is demonstrated in Fig.~\ref{fig:Fig2}c and Fig.~\ref{fig:Fig2}d. For example, if a photon is injected into port 1 for the state $|000\rangle$, then a unitary of $\sigma_1^0\otimes\sigma_2^0\otimes\sigma_3^x$ can route the photon to port 2 to get a state $|001\rangle$, where qubit 3 is flipped. The global encoding, as a product of codesign, has the ability of directly implementing the whole unitary, taking full advantage of our processor. This advantage, due to the different evolution formulations of the CCQO and the CCQO-E, is more prominent in CCQO-E.

\begin{figure*}[t!]
\includegraphics[width=0.93\textwidth]{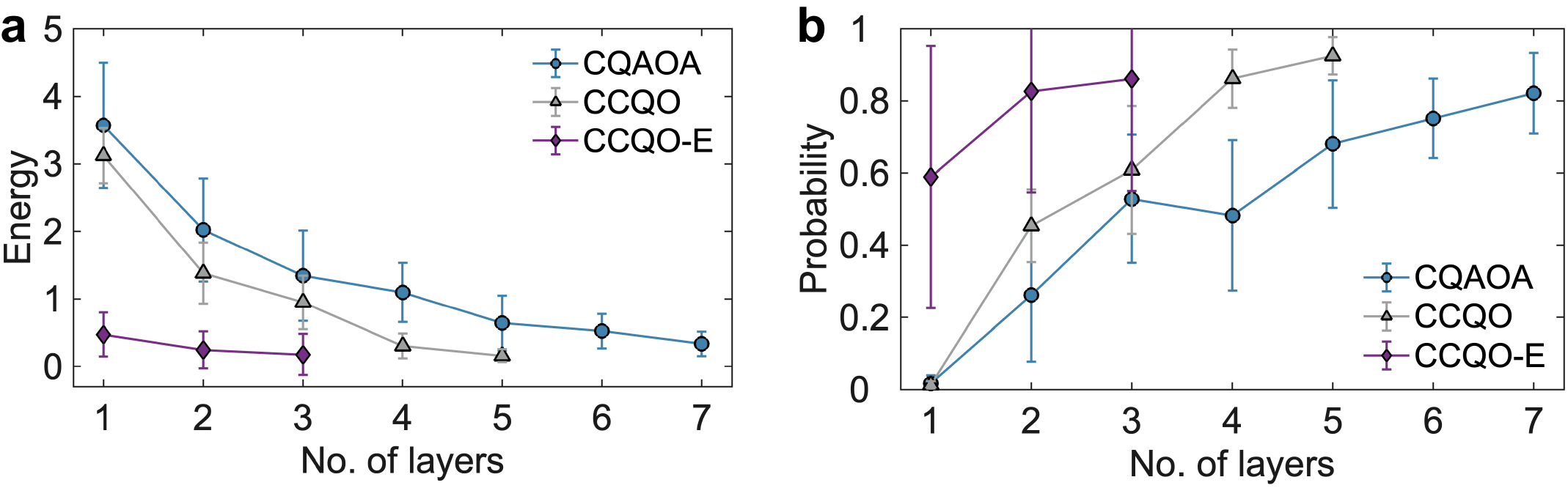}
\caption{\textbf{Experimental results of factoring the number 2893 using the three algorithms as a function of number of layers.} {\bf a } Experimentally measured energy at the end of the iteration. {\bf b} Experimentally measured success probability  of obtaining the ground state $|011\rangle$ at the end of the iteration. In each subgraph, The blue, grey and purple lines are respectively the results of CQAOA, CCQO and CCQO-E. The dots and error bars are respectively the average and the standard deviations of 10 experimental results. In each of the 10 experiments, the ansatz is initialized randomly with all parameters chosen in the range of $[-\pi,\pi)$.}
\label{fig:Fig4}
\end{figure*}

To compare the structures of CQAOA, CCQO and CCQO-E, we illustrate the quantum circuit by unitary blocks, as shown in Fig.~\ref{fig:Fig2}e. The ansatz is initialized randomly and loaded to the PIC, where the 8-mode PIC is mapped to a three-qubit circuit. The light intensity distribution in each mode can be regarded as an average measurement result on different computational bases. Thus the expectation value $\langle\psi_{\mathrm{in}}|U(\alpha,\beta,\gamma)^{\dagger}H_{\mathrm{p}}U(\alpha,\beta,\gamma)|\psi_{\mathrm{in}}\rangle$, $i.e.$, the energy of the problem Hamiltonian in CCQO for example, can be evaluated and used as the cost function in the following optimization process.  The gradient is obtained by finite difference method, calculated with two forward propagating whose parameter have a slight shift, e.g. 
${\partial E}/{\partial\alpha}=[\langle\psi_{\mathrm{in}}|U(\alpha+\delta\alpha,\beta,\gamma)^{\dagger}H_{\mathrm{p}}U(\alpha,\beta,\gamma)|\psi_{\mathrm{in}}\rangle-\langle\psi_{\mathrm{in}}|U(\alpha-\delta\alpha,\beta,\gamma)^{\dagger}H_{\mathrm{p}}U(\alpha,\beta,\gamma)|\psi_{\mathrm{in}}\rangle]/2\delta\alpha$.    
Note that depending on the choice of algorithms, CQAOA, CCQO, and CCQO-E hold different parameters in one layer and the layer is repeated $p$ times to exhibit distinct optimization effects.
The learning rate is initially set to 0.005 and rescaled to 10\% of its previous value every 30 iterations.

\subsection{Algorithm performance}

Specifically, we try to benchmark the CCQO-E by factorizing the number 2893 via a variational approach on a classical-quantum processor. To map the prime factorization problem on a photonic circuit, we first introduce the problem Hamiltonian through the binary multiplication table \cite{burges2002factoring}, where two prime factors are respectively represented as binary numbers $p = (1\ x_7\ x_6\ x_5\ x_4\ x_3\ x_2\ x_1\ 1)$ and $q = (1\ y_2\ y_1\ 1)$. The first and last bits are set to 1 to ensure two odd numbers. The factorization equations are obtained by multiplying $p$ and $q$ and adding each column of the multiplication table. Since the number of qubits is fixed, we simplify these equations further to reduce the total qubit requirement by applying classical prepossessing based on binary logical constraints. Mapping the variables in the objective function (see Methods) to qubit operators gives the problem Hamiltonian, 
\begin{eqnarray}\label{eq2}
H_{\mathrm{p}}=&0.75\sigma_1^z \sigma_2^z \sigma_3^z + 2\sigma_1^z \sigma_2^z - 0.25\sigma_1^z \sigma_3^z \nonumber \\
&- 1.75\sigma_2^z \sigma_3^z - 1.5\sigma_1^z - 2\sigma_2^z + 1.75\sigma_3^z,
\end{eqnarray}
where the constant term $4I$ is dropped. The ground state of this Hamiltonian 
$|011\rangle$ encodes the solution to the factorization problem.  To obtain the ground state, we consider the CCQO-E, which is the optimized version of CCQO. 

The main advantage of the ansatz in CCQO-E is that it does not require implementing the many-body terms in the problem Hamiltonian. Additionally, the circuit ansatz is problem-inspired and consists only of local CD terms selected from the operator pool. The number of parameters scales with the number of operators chosen from the pool. Although we chose the operators randomly, one could also rely on gradient-based selection of operators, as in the adaptive variational algorithm \cite{grimsley2019adaptive}.

In the experiment, the ansatz of the three algorithms is separately mapped onto the photonic circuit by calculating and decomposing the matrix in a classical computer. These ansatz matrices are determined by the circuit parameters: QAOA utilizes only the problem Hamiltonian and the mixing Hamiltonian, resulting in 2 parameters ($\beta$ and $\gamma$ in Fig.~\ref{fig:Fig2}e) per layer. In CCQO, the inclusion of CD terms increases this to 3 parameters ($\alpha$, $\beta$ and $\gamma$ in Fig.~\ref{fig:Fig2}e) per layer. In the problem-inspired ansatz of CCQO-E, there are 3 distinct parameters, $\alpha_{1,2,3}$, corresponding to local interactions, and 1 parameter, $\beta$, corresponding to two-body interactions.
The initial state is set to be $|000\rangle$, $i.e.$ the light is injected into the first port and the parameters in the ansatz are initialized randomly in the range of $[-\pi,\pi)$. Each experiment is repeated 10 times with different initial parameters. 

The optimization processes of 1-layer, 2-layer and 3-layer circuits are depicted in Fig.~\ref{fig:Fig3}, showing the energy descent and success probability ascent curves during iteration. For the problem Hamiltonian of factorizing 2893, the ground state is $|110\rangle$ with a corresponding ground energy of 0.
In the experiment, we measure the light intensity of each of the eight ports. Each light intensity is normalized by the total intensity of the eight ports to obtain the probability. During iterations, we measure the light intensity after each step, as shown in Fig.~\ref{fig:Fig3}a-c. Fig.~\ref{fig:Fig3}a, b and c are respectively the normalized light intensity of CQAOA, CCQO and CCQO-E with 3 layers. Each color corresponds to a specific quantum state, as displayed in the legend. The eight color blocks are stacked together to identity. The area of the blue block is getting larger with the number of steps, indicating the state $|011\rangle$ is found. Comparing Fig.\ref{fig:Fig3}a, b, and c, we find the CCQO-E is most probable to find the target solution. 

In addition to the single experiment shown in Fig.~\ref{fig:Fig3}a-c, we randomly sample the initial ansatz and perform 10 different experiments for each algorithm and each number of layers. In Fig.~\ref{fig:Fig3}d-f, the purple, blue and red lines are respectively the average energy $\bar{\epsilon}$ of CQAOA, CCQO and CCQO-E. The dashed lines are the corresponding standard deviations $\Delta\epsilon$. For all there algorithms, the final energy decreases as the number of layers increases, indicating that increasing the number of layers contributes to the proximity to the ground state.
When the circuit has 3 layers, the average energies after iterations for CQAOA, CCQO and CCQO-E are 3.57, 3.13 and 0.472, respectively, as shown in Fig.~\ref{fig:Fig3}f. 
In Fig.~\ref{fig:Fig3}d-f, we also observe a clear advantage with CCQO-E. In the cases of 1 layer, 2 layers and 3 layers, the red line is always steeper than the purple and blue lines in the beginning of iterations, which means a faster search for the ground state. 
Specifically, using the optimized algorithm, the system energy converges below 0.5 after approximately 100 iterations even in a 1-layer circuit. This is lower energy than the 3-layer CQAOA and the 3-layer common CCQO to converge. 
Considering the number of parameters used in 1-layer CCQO-E and 3-layer CCQO and CQAOA are respectively 4, $3\times3$ and $2\times3$, we find the optimized algorithm is resource efficient. 

We also compare the success probability, $i.e.$, the overlap between the trial state after optimization and $|110\rangle$. In Fig.~\ref{fig:Fig3}g-i, the purple, blue and red lines are respectively the average success probability $\bar{P}$ of CQAOA, CCQO and CCQO-E. The dashed lines are the corresponding standard deviations $\Delta P$.
It can be seen that the success probability increases with the number of layers. The average success probabilities of CQAOA, CCQO and CCQO-E with 3 layers are 52.8\%, 60.8\% and 86.1\%, respectively, as shown in Fig.~\ref{fig:Fig3}i. 
We find that for each number of layers the red line is much higher than the purple and blue lines, and the success probability of 1-layer CCQO-E is approximately equal to that of 3-layer CCQO. This, again, demonstrates the optimized algorithm is resource friendly. 

The introduction of CD driving not only produces a more accurate trial state for finding the target ground state, but also reduces the circuit depth and decreases the number of parameters needed to achieve comparable performance. Therefore, we evaluate the performance of the series of quantum optimization algorithms across various circuit depths, as shown in Fig.~\ref{fig:Fig4}. The CCQO-E demonstrates significant optimization improvements even when using a shallow circuit with $p=1$. In contrast, both CQAOA and CCQO fail to perform valid optimization at this circuit depth. Furthermore, CCQO-E achieves near-optimal results with just $p=3$ layers, whereas CQAOA and CCQO require 7 and 5 layers, respectively, to reach similar performances.

\section*{Discussion}
In this work, we present an experimental realization of codesigned variational quantum algorithms on our universal photonic integrated circuit. The global encoding approach tailors the algorithms for seamless integration with photonic processors, which would not be made possible using the typical dual-rail encoding approach. The latter faces inherent challenges such as the probabilistic CNOT gate and limited circuit depth, and in the meantime, wastes a lot of available manipulable parameters throughout the dense MZI mesh. A codesign via a more compatible encoding strategy breaks through the challenges of executing complex quantum circuits on photonic systems while fully leveraging the strengths of the photonic technology.

In particular, we experimentally demonstrate an enhanced codesigned counterdiabatic quantum optimization algorithm. By mapping the factorization of 2893 to a ground state searching problem, we evaluate the performance of CCQO-E in comparison with CQAOA and CCQO. Notably, CCQO-E achieves a better trial state that more closely matches the ground state of the problem Hamiltonian compared to its predecessors with the same circuit depths. Additionally, this method significantly reduces the number of parameters required to solve the same task, as demonstrated by its high performance in our experimental results, even with shallow circuit depths. It is important to note that these results provide an impartial benchmark of a series of algorithms, which, in other systems, are often affected by differences in gate fidelity and circuit depth.


We propose that codesign would play an ever growing role in quantum information technologies. Whilst the `hardware-agnostic' quantum algorithms are being driven by the fault-tolerant quantum computing researches, the codesign that exploits the maximal of NISQ device capabilities would still be of a central concept. Recent works, such as digital-analog hybrid quantum simulation~\cite{parra2020digital,martin2023digital,Tao2021digital}, are other good examples of quantum codesign. Furthermore, the codesign technique can be applied to photonic devices in various ways beyond customizing the encoding method. For instance, universal MZI meshes can be partitioned into blocks, cascaded in groups, or organized in other configurations inspired by the specific problem to be solved. PICs fabricated with advanced materials such as SiN, LNOI, and SiC offer significant advantages in optical efficiency, high bandwidth, and ultra-low energy consumption, enable an optimal synergy between classical and quantum computational resources, optimizing the performance of hybrid classical-quantum systems. Our work paves the way for the advancement of photonic processors in quantum computing and quantum information processing. 

\emph{Notes}: We get aware of a related but not overlapping work \cite{Agresti2024} recently.

\section*{Methods}

\subsection*{Construction of Hamiltonians for CCQO }

{
The common algorithms, like QAOA, requires quantum circuits of considerable depth, beyond the capabilities of near-term quantum computers. To address this challenge, we consider the CD protocol to tackle the optimization problem, where the CD driving accelerates adiabatic evolution by compensating for non-adiabatic transitions that arise during fast evolution \cite{hegade2022digitized}. Moreover, to address the challenge that the exact CD terms are computationally expensive and challenging, the nested commutator method has been proposed to provide a practical way to approximate the CD term \cite{chandarana2023digitized}. This approach simplifies implementation while maintaining high fidelity, making it more suitable for near-term quantum devices. }

{
Consider an adiabatic evolution described by a time-dependent scheduling parameter \( \lambda(t) \), which governs the system’s evolution from the initial to the final state, where the Hamiltonian is changing from initial Hamiltonian $H_{\mathrm{i}}$ to final Hamiltonian $H_{\mathrm{f}}$. Under the CD protocol, the total Hamiltonian is modified by an additional counterdiabatic term, involving the adiabatic gauge potential \( A_{\lambda} \), which compensates for non-adiabatic transitions. The counterdiabatic term is given by \( \dot{\lambda} A_{\lambda} \), where \( \dot{\lambda} \) is the rate of change of the scheduling parameter. For many-body systems, obtaining the exact \( A_{\lambda} \) is challenging due to its complexity. To overcome this, the adiabatic gauge potential can be approximated using a nested commutator expansion, where \( A_{\lambda} \) is expressed as a series of commutators involving the system's Hamiltonian and its derivative with respect to \( \lambda \):
\begin{equation}
    A_{\lambda} = \i \sum_{l} \alpha_l \left[ H_{\mathrm{ad}}(\lambda), \left[ H_{\mathrm{ad}}(\lambda), \dots, \left[ H_{\mathrm{ad}}(\lambda), \partial_{\lambda} H_{\mathrm{ad}}(\lambda) \right] \dots \right] \right] .
\end{equation}
The coefficients \( \alpha_l \), known as CD coefficients, are determined by variational principles by minimizing the action \cite{sels2017minimizing}. Here, $H_{\mathrm{ad}}$ is the adiabatic Hamiltonian 
\begin{equation}
    H_{\mathrm{ad}} = \left[ 1-\lambda(t) \right] H_{\mathrm{i}} + \lambda(t) H_{\mathrm{f}} .
\end{equation}
Our work adapts this nested commutators to achieve the on-chip CCQO and CCQO-E. For the problem Hamiltonian in Eq. (\ref{eq2}), we define an operator pool containing local CD terms obtained from the lower-order nested commutator ansatz, $A = \{\sigma_y, \sigma_z \sigma_y, \sigma_y \sigma_z , \sigma_x \sigma_y , \sigma_y \sigma_x \}$.
}

The objective function after classical preprocessing is given by
\begin{align}\label{eq1}
    f(x,y,c)= & -6x_6 y_1 c_5 +2x_6 c_5 -4y_1 c_5 -c_5 \nonumber \\
    &+11x_6 y_1 -2x_6 +2y_1 +3 .
\end{align}
where $c_i$'s are binary variables representing the carriers. Mapping the variables to qubit operator $k_i\rightarrow(1-\sigma_i^z)/2$ gives the problem Hamiltonian in Eq.~(\ref{eq2}). Since the operator pool is $A = \{\sigma_y, \sigma_z \sigma_y, \sigma_y \sigma_z , \sigma_x \sigma_y , \sigma_y \sigma_x \}$, we choose the CD Hamiltonian in our experiments to be:
\begin{align}\label{eq3}
    H_{\mathrm{CD}} = & 2(\sigma_1^z \sigma_2^y +\sigma_1^y \sigma_2^z)-0.25(\sigma_1^z \sigma_3^y +\sigma_1^y \sigma_3^z) \nonumber \\
    &-1.75(\sigma_2^z \sigma_3^y +\sigma_2^y \sigma_3^z) .
\end{align}
The mixing term is chosen as $\sigma^x$, giving the mixing Hamiltonian:
\begin{equation}
    H_\mathrm{m} = - \sigma_1^x - \sigma_2^x - \sigma_3^x.
\end{equation}

\subsection*{Experimental set-up}

The experimental illustration of CCQO-E in this work is conducted with a PIC and a personal computer. The PIC is composed of 36 cascaded MZIs. According to Clements \cite{clements2016optimal}, aside from the interferometers upper and lower most in even columns which are used to compensate the optical path difference, the rest MZI mesh can perform 8-mode arbitrary unitary transformation. Light in waveguides is modulated by two thermo-optic PS and two MMIs. The PSs provide a tunable phase shift by applying an appropriate current controlled by an FPGA. We denote the internal (external) phase shift as $2\theta$ ($\phi$). The MMIs are functioned as 50:50 beam splitters where light from two arms interferences. The characterization method of PSs can be found in SM. 

The PIC, FPGA, power detector, and the cooling fan are integrated on a development board
. Light from a tunable continuous wave laser is modulated by a fibre polarization controller and then injected to a fibre array. The intensity of the output light is measured using on-chip power detectors, and the information exchange between the PIC and the personal computer is established via a LAN port.
PSs are heated by electronodes linked to the PCB board with gold wires and the input (output) light from (to) FA is coupled in (out) the PIC with grating structures.




\section*{Acknowledgments} 
The authors thank Hai-Rui Zhang, Ke-Ming Hu and Feng-Kai Han for helpful discussions. The computations in this paper were run on the $\pi$ 2.0 cluster supported by the Center for High Performance Computing at Shanghai Jiao Tong University. 
This research was supported by the National Key R\&D Program of China (Grant No. 2019YFA0706302, No. 2019YFA0308703, No. 2017YFA0303700); the National Natural Science Foundation of China (Grants No. 62235012, No. 11904299, No. 61734005, No. 11761141014, and No. 11690033, No. 12104299, and No. 12304342); Innovation Program for Quantum Science and Technology (Grants No. 2021ZD0301500, and No. 2021ZD0300700); Science and Technology Commission of Shanghai Municipality (STCSM) (Grants No. 20JC1416300, No. 2019SHZDZX01, No. 21ZR1432800, No. 22QA1404600), and the Shanghai Municipal Education Commission (SMEC) (2017-01-07-00-02-E00049). China Postdoctoral Science Foundation (Grants No. 2020M671091, No. 2021M692094, No. 2022T150415). X.-M.J. acknowledges additional support from a Shanghai talent program and support from Zhiyuan Innovative Research Center of Shanghai Jiao Tong University. H. T. acknowledges additional support from Yangyang Development Fund.

\section*{Author contributions}
X.-W.S. and H.T. designed the experiments.
X.C., Z.-F.L. and X.-W.S. performed the experiments.
N.N.H. built the theoretical model. 
X.-W.S. and N.N.H. conducted numerical simulations. 
X.-W.S. and X.C. analyzed the experimental data.
X.-K.L. and Y.-Q.P. fabricated the device. 
X.-W.S., X.C., N.N.H., H.T. and E.S. contributed to the manuscript. 
H.T., E.S. and X.-M.J. supervised the project.

\section*{Competing interests}
The authors declare no competing interests.


%





\begin{thebibliography}{99}

\bibitem{barends2016digitized}
\bibinfo{author}{Barends, R. \textit{et al.} }
\bibinfo{title}{Digitized adiabatic quantum computing with a superconducting circuit. }
\newblock \emph{\bibinfo{journal}{Nature}}
\textbf{\bibinfo{volume}{534}}, \bibinfo{pages}{222-226}
  (\bibinfo{year}{2016}).


\bibitem{wendin2017quantum}
\bibinfo{author}{Wendin, G. }
\bibinfo{title}{Quantum information processing with superconducting circuits: a review. }
\newblock \emph{\bibinfo{journal}{Nature}}
\textbf{\bibinfo{volume}{80}}, \bibinfo{pages}{106001}
(\bibinfo{year}{2016}).


\bibitem{xbxuchip2022}
\bibinfo{author}{Xu, X.-B. \textit{et al.} }
\bibinfo{title}{Hybrid superconducting photonic-phononic chip for quantum information processing.}
\newblock \emph{\bibinfo{journal}{Chip}} \textbf{\bibinfo{volume}{1}}, \bibinfo{pages}{100016} (\bibinfo{year}{2022}).


\bibitem{leibfried2003quantum}
\bibinfo{author}{Leibfried, D.}, \bibinfo{author}{Blatt, R.}, \bibinfo{author}{Monroe, C.} \& \bibinfo{author}{Wineland, D.}
\bibinfo{title}{Quantum dynamics of single trapped ions.}
\newblock \emph{\bibinfo{journal}{Reviews of Modern Physics}} \textbf{\bibinfo{volume}{75}}, \bibinfo{pages}{281} (\bibinfo{year}{2003}).

\bibitem{blatt2012quantum}
\bibinfo{author}{Blatt, R.} \& \bibinfo{author}{Roos, C. F.}
\bibinfo{title}{Quantum simulations with trapped ions.}
\newblock \emph{\bibinfo{journal}{Nature Physics}} \textbf{\bibinfo{volume}{8}}, \bibinfo{pages}{277-284} (\bibinfo{year}{2012}).

\bibitem{carolan2015universal}
\bibinfo{author}{Carolan, J.} \textit{et al.}
\bibinfo{title}{Universal linear optics.}
\newblock \emph{\bibinfo{journal}{Science}} \textbf{\bibinfo{volume}{349}}, \bibinfo{pages}{711-716} (\bibinfo{year}{2015}).

\bibitem{tillmann2013experimental}
\bibinfo{author}{Tillmann, M.} \textit{et al.}
\bibinfo{title}{Experimental boson sampling.}
\newblock \emph{\bibinfo{journal}{Nature Photonics}} \textbf{\bibinfo{volume}{7}}, \bibinfo{pages}{540-544} (\bibinfo{year}{2013}).

\bibitem{daley2022practical}
\bibinfo{author}{Daley, A. J.} \textit{et al.}
\bibinfo{title}{Practical quantum advantage in quantum simulation.}
\newblock \emph{\bibinfo{journal}{Nature}} \textbf{\bibinfo{volume}{607}}, \bibinfo{pages}{667-676} (\bibinfo{year}{2022}).

\bibitem{huang2022quantum}
\bibinfo{author}{Huang, H.-Y.} \textit{et al.}
\bibinfo{title}{Quantum advantage in learning from experiments.}
\newblock \emph{\bibinfo{journal}{Science}} \textbf{\bibinfo{volume}{376}}, \bibinfo{pages}{1182-1186} (\bibinfo{year}{2022}).

\bibitem{zhong2020quantum}
\bibinfo{author}{Zhong, H.-S.} \textit{et al.}
\bibinfo{title}{Quantum computational advantage using photons.}
\newblock \emph{\bibinfo{journal}{Science}} \textbf{\bibinfo{volume}{370}}, \bibinfo{pages}{1460-1463} (\bibinfo{year}{2020}).

\bibitem{wu2021strong}
\bibinfo{author}{Wu, Y.} \textit{et al.}
\bibinfo{title}{Strong quantum computational advantage using a superconducting quantum processor.}
\newblock \emph{\bibinfo{journal}{Physical Review Letters}} \textbf{\bibinfo{volume}{127}}, \bibinfo{pages}{180501} (\bibinfo{year}{2021}).


\bibitem{grover1996fast}
\bibinfo{author}{Grover, L. K.}
\bibinfo{title}{A fast quantum mechanical algorithm for database search.}
\newblock In \emph{\bibinfo{booktitle}{Proceedings of the twenty-eighth annual ACM symposium on Theory of computing}}, \bibinfo{pages}{212-219} (1996).

\bibitem{shor1994algorithms}
\bibinfo{author}{Shor, P. W.}
\bibinfo{title}{Algorithms for quantum computation: discrete logarithms and factoring.}
\newblock In \emph{\bibinfo{booktitle}{Proceedings 35th annual symposium on foundations of computer science}}, \bibinfo{pages}{124-134} (1994).

\bibitem{Schuld2015}
\bibinfo{author}{Schuld, M.} \textit{et al.}
\bibinfo{title}{An introduction to quantum machine learning.}
\newblock \emph{\bibinfo{journal}{Contemporary Physics}} \textbf{\bibinfo{volume}{56}}, \bibinfo{pages}{172-185} (2015).

\bibitem{fowler2012surface}
\bibinfo{author}{Fowler, A. G.} \textit{et al.}
\bibinfo{title}{Surface codes: Towards practical large-scale quantum computation.}
\newblock \emph{\bibinfo{journal}{Physical Review A}} \textbf{\bibinfo{volume}{86}}, \bibinfo{pages}{032324} (2012).

\bibitem{konno2024logical}
\bibinfo{author}{Konno, S.} \textit{et al.}
\bibinfo{title}{Logical states for fault-tolerant quantum computation with propagating light.}
\newblock \emph{\bibinfo{journal}{Science}} \textbf{\bibinfo{volume}{383}}, \bibinfo{pages}{289-293} (2024).

\bibitem{michael2016new}
\bibinfo{author}{Michael, M. H.} \textit{et al.}
\bibinfo{title}{New class of quantum error-correcting codes for a bosonic mode.}
\newblock \emph{\bibinfo{journal}{Physical Review X}} \textbf{\bibinfo{volume}{6}}, \bibinfo{pages}{031006} (2016).

\bibitem{cai2023quantum}
\bibinfo{author}{Cai, Z.} \textit{et al.}
\bibinfo{title}{Quantum error mitigation.}
\newblock \emph{\bibinfo{journal}{Reviews of Modern Physics}} \textbf{\bibinfo{volume}{95}}, \bibinfo{pages}{045005} (2023).

\bibitem{Gebhart2023}
\bibinfo{author}{Gebhart, V.} \textit{et al.}
\bibinfo{title}{Learning quantum systems.}
\newblock \emph{\bibinfo{journal}{Nature Reviews Physics}} \textbf{\bibinfo{volume}{5}}, \bibinfo{pages}{141-156} (2023).

\bibitem{bharti2022noisy}
\bibinfo{author}{Bharti, K.} \textit{et al.}
\bibinfo{title}{Noisy intermediate-scale quantum algorithms.}
\newblock \emph{\bibinfo{journal}{Reviews of Modern Physics}} \textbf{\bibinfo{volume}{94}}, \bibinfo{pages}{015004} (2022).

\bibitem{hartmann2016quantum}
\bibinfo{author}{Hartmann, M. J.}
\bibinfo{title}{Quantum simulation with interacting photons.}
\newblock \emph{\bibinfo{journal}{Journal of Optics}} \textbf{\bibinfo{volume}{18}}, \bibinfo{pages}{104005} (2016).

\bibitem{cerezo2021variational}
\bibinfo{author}{Cerezo, M.} \textit{et al.}
\bibinfo{title}{Variational quantum algorithms.}
\newblock \emph{\bibinfo{journal}{Nature Reviews Physics}} \textbf{\bibinfo{volume}{3}}, \bibinfo{pages}{625-644} (2021).

\bibitem{benedetti2021hardware}
\bibinfo{author}{Benedetti, M.} \textit{et al.}
\bibinfo{title}{Hardware-efficient variational quantum algorithms for time evolution.}
\newblock \emph{\bibinfo{journal}{Physical Review Research}} \textbf{\bibinfo{volume}{3}}, \bibinfo{pages}{033083} (2021).

\bibitem{lubasch2020variational}
\bibinfo{author}{Lubasch, M.} \textit{et al.}
\bibinfo{title}{Variational quantum algorithms for nonlinear problems.}
\newblock \emph{\bibinfo{journal}{Physical Review A}} \textbf{\bibinfo{volume}{101}}, \bibinfo{pages}{010301} (2020).

\bibitem{farhi2014quantum}
\bibinfo{author}{Farhi, E.} \textit{et al.}
\bibinfo{title}{A quantum approximate optimization algorithm.}
\newblock \emph{\bibinfo{journal}{arXiv preprint}} arXiv:1411.4028 (2014).

\bibitem{vikstaal2020applying}
\bibinfo{author}{Vikst{\aa}l, P.} \textit{et al.}
\bibinfo{title}{Applying the quantum approximate optimization algorithm to the tail-assignment problem.}
\newblock \emph{\bibinfo{journal}{Physical Review Applied}} \textbf{\bibinfo{volume}{14}}, \bibinfo{pages}{034009} (2020).

\bibitem{hodson2019portfolio}
\bibinfo{author}{Hodson, M.} \textit{et al.}
\bibinfo{title}{Portfolio rebalancing experiments using the quantum alternating operator ansatz.}
\newblock \emph{\bibinfo{journal}{arXiv preprint}} arXiv:1911.05296 (2019).

\bibitem{brady2021optimal}
\bibinfo{author}{Brady, L. T.} \textit{et al.}
\bibinfo{title}{Optimal protocols in quantum annealing and quantum approximate optimization algorithm problems.}
\newblock \emph{\bibinfo{journal}{Physical Review Letters}} \textbf{\bibinfo{volume}{126}}, \bibinfo{pages}{070505} (2021).

\bibitem{zhu2022adaptive}
\bibinfo{author}{Zhu, L.} \textit{et al.}
\bibinfo{title}{Adaptive quantum approximate optimization algorithm for solving combinatorial problems on a quantum computer.}
\newblock \emph{\bibinfo{journal}{Physical Review Research}} \textbf{\bibinfo{volume}{4}}, \bibinfo{pages}{033029} (2022).

\bibitem{bravyi2020obstacles}
\bibinfo{author}{Bravyi, S.} \textit{et al.}
\bibinfo{title}{Obstacles to variational quantum optimization from symmetry protection.}
\newblock \emph{\bibinfo{journal}{Physical Review Letters}} \textbf{\bibinfo{volume}{125}}, \bibinfo{pages}{260505} (2020).

\bibitem{egger2021warm}
\bibinfo{author}{Egger, D. J.} \textit{et al.}
\bibinfo{title}{Warm-starting quantum optimization.}
\newblock \emph{\bibinfo{journal}{Quantum}} \textbf{\bibinfo{volume}{5}}, \bibinfo{pages}{479} (2021).

\bibitem{farhi2000quantum}
\bibinfo{author}{Farhi, E.} \textit{ et al.} \bibinfo{title}{Quantum computation by adiabatic evolution.}
\textit{arXiv preprint} arXiv:quant-ph/0001106 (2000).

\bibitem{demirplak2003adiabatic}
\bibinfo{author}{Demirplak, M. and Rice, S. A.}
\bibinfo{title}{Adiabatic population transfer with control fields.} \newblock \emph{\bibinfo{journal}{The Journal of Physical Chemistry A}} \textbf{\bibinfo{volume}{107}}, \bibinfo{pages}{9937-9945} (2003).

\bibitem{berry2009transitionless}
\bibinfo{author}{Berry, M. V.}
\bibinfo{title}{Transitionless quantum driving.}
\newblock \emph{\bibinfo{journal}{Journal of Physics A: Mathematical and Theoretical}} \textbf{\bibinfo{volume}{42}}, \bibinfo{pages}{365303} (2009).

\bibitem{del2013shortcuts}
\bibinfo{author}{Del Campo, A.}
\bibinfo{title}{Shortcuts to adiabaticity by counterdiabatic driving.}
\newblock \emph{\bibinfo{journal}{Physical Review Letters}} \textbf{\bibinfo{volume}{111}}, \bibinfo{pages}{100502} (2013).

\bibitem{chandarana2022digitized}
\bibinfo{author}{Chandarana, P.} \textit{et al.}
\bibinfo{title}{Digitized-counterdiabatic quantum approximate optimization algorithm.}
\newblock \emph{\bibinfo{journal}{Physical Review Research}} \textbf{\bibinfo{volume}{4}}, \bibinfo{pages}{013141} (2022).

\bibitem{hegade2022digitized}
\bibinfo{author}{Hegade, N. N.} \textit{et al.}
\bibinfo{title}{Digitized counterdiabatic quantum optimization.}
\newblock \emph{\bibinfo{journal}{Physical Review Research}} \textbf{\bibinfo{volume}{4}}, \bibinfo{pages}{L042030} (2022).

\bibitem{sels2017minimizing}
\bibinfo{author}{Sels, D.} \textit{et al.}
\bibinfo{title}{Minimizing irreversible losses in quantum systems by local counterdiabatic driving.}
\newblock \emph{\bibinfo{journal}{Proceedings of the National Academy of Sciences}} \textbf{\bibinfo{volume}{114}}, \bibinfo{pages}{E3909--E3916} (2017).

\bibitem{claeys2019floquet}
\bibinfo{author}{Claeys, P. W.} \textit{et al.}
\bibinfo{title}{Floquet-engineering counterdiabatic protocols in quantum many-body systems.}
\newblock \emph{\bibinfo{journal}{Physical Review Letters}} \textbf{\bibinfo{volume}{123}}, \bibinfo{pages}{090602} (2019).

\bibitem{peruzzo2014variational}
\bibinfo{author}{Peruzzo, A.} \textit{et al.}
\bibinfo{title}{A variational eigenvalue solver on a photonic quantum processor.}
\newblock \emph{\bibinfo{journal}{Nature Communications}} \textbf{\bibinfo{volume}{5}}, \bibinfo{pages}{4213} (2014).

\bibitem{chen2020demonstration}
\bibinfo{author}{Chen, M.-C.} \textit{et al.}
\bibinfo{title}{Demonstration of adiabatic variational quantum computing with a superconducting quantum coprocessor.}
\newblock \emph{\bibinfo{journal}{Physical Review Letters}} \textbf{\bibinfo{volume}{125}}, \bibinfo{pages}{180501} (2020).

\bibitem{hempel2018quantum}
\bibinfo{author}{Hempel, C.} \textit{et al.}
\bibinfo{title}{Quantum chemistry calculations on a trapped-ion quantum simulator.}
\newblock \emph{\bibinfo{journal}{Physical Review X}} \textbf{\bibinfo{volume}{8}}, \bibinfo{pages}{031022} (2018).

\bibitem{ebadi2022quantum}
\bibinfo{author}{Ebadi, S.} \textit{et al.}
\bibinfo{title}{Quantum optimization of maximum independent set using Rydberg atom arrays.}
\newblock \emph{\bibinfo{journal}{Science}} \textbf{\bibinfo{volume}{376}}, \bibinfo{pages}{1209-1215} (2022).

\bibitem{xue2022variational}
\bibinfo{author}{Xue, S.} \textit{et al.}
\bibinfo{title}{Variational entanglement-assisted quantum process tomography with arbitrary ancillary qubits.}
\newblock \emph{\bibinfo{journal}{Physical Review Letters}} \textbf{\bibinfo{volume}{129}}, \bibinfo{pages}{133601} (2022).

\bibitem{carolan2020variational}
\bibinfo{author}{Carolan, J.} \textit{et al.}
\bibinfo{title}{Variational quantum unsampling on a quantum photonic processor.}
\newblock \emph{\bibinfo{journal}{Nature Physics}} \textbf{\bibinfo{volume}{16}}, \bibinfo{pages}{322-327} (2020).

\bibitem{cimini2024variational}
\bibinfo{author}{Cimini, V.} \textit{et al.}
\bibinfo{title}{Variational quantum algorithm for experimental photonic multiparameter estimation.}
\newblock \emph{\bibinfo{journal}{npj Quantum Information}} \textbf{\bibinfo{volume}{10}}, \bibinfo{pages}{26} (2024).

\bibitem{dnliuchip2022}
\bibinfo{author}{Liu, D.-N.} \textit{et al.}
\bibinfo{title}{Generation and dynamic manipulation of frequency degenerate polarization entangled Bell states by a silicon quantum photonic circuit.}
\newblock \emph{\bibinfo{journal}{Chip}} \textbf{\bibinfo{volume}{1}}, \bibinfo{pages}{100001} (2022).

\bibitem{xdwangchip2022}
\bibinfo{author}{Wang, X.-D.} \textit{et al.}
\bibinfo{title}{Waveguide-coupled deterministic quantum light sources and post-growth engineering methods for integrated quantum photonics.}
\newblock \emph{\bibinfo{journal}{Chip}} \textbf{\bibinfo{volume}{1}}, \bibinfo{pages}{100018} (2022).

\bibitem{Somhorst2203}
\bibinfo{author}{Somhorst, F. H. B.} \textit{et al.}
\bibinfo{title}{Quantum simulation of thermodynamics in an integrated quantum photonic processor.}
\newblock \emph{\bibinfo{journal}{Nature Communications}} \textbf{\bibinfo{volume}{14}}, \bibinfo{pages}{3895} (2023).


\bibitem{li2024high}
\bibinfo{author}{Li, X.-K.} \textit{et al.}
\bibinfo{title}{High-efficiency reinforcement learning with hybrid architecture photonic integrated circuit.}
\newblock \emph{\bibinfo{journal}{Nature Communications}} \textbf{\bibinfo{volume}{15}},  \bibinfo{pages}{1044} (2024).

\bibitem{tomesh2021quantum}
\bibinfo{author}{Tomesh, T.} \& \bibinfo{author}{Martonosi, M.}
\bibinfo{title}{Quantum codesign.}
\newblock \emph{\bibinfo{journal}{IEEE Micro}} \textbf{\bibinfo{volume}{41}}, \bibinfo{pages}{33--40} (2021).

\bibitem{burges2002factoring}
\bibinfo{author}{Burges, C. J. C.}
\bibinfo{title}{Factoring as optimization.}
\newblock \emph{\bibinfo{journal}{Microsoft Research MSR-TR-200}} (2002).

\bibitem{grimsley2019adaptive}
\bibinfo{author}{Grimsley, H. R.} \textit{et al.}
\bibinfo{title}{An adaptive variational algorithm for exact molecular simulations on a quantum computer.}
\newblock \emph{\bibinfo{journal}{Nature Communications}} \textbf{\bibinfo{volume}{10}}, \bibinfo{pages}{3007} (2019).

\bibitem{parra2020digital}
\bibinfo{author}{Parra, R.}\textit{et al.}
\bibinfo{title}{Digital-analog quantum computation.}
\newblock \emph{\bibinfo{journal}{Physical Review A}} \textbf{\bibinfo{volume}{101}}, \bibinfo{pages}{022305} (2020).

\bibitem{martin2023digital}
\bibinfo{author}{Martin, A., Ibarrondo, R., \& Sanz, M.}
\bibinfo{title}{Digital-analog co-design of the Harrow-Hassidim-Lloyd algorithm}
\newblock \emph{\bibinfo{journal}{Physical Review A}} \textbf{\bibinfo{volume}{19}}, \bibinfo{pages}{064056} (2023).

\bibitem{Tao2021digital}
\bibinfo{author}{Tao, Z.}\textit{et al.}
\bibinfo{title}{Experimental realization of phase-controlled dynamics with hybrid digital-analog approach.}
\newblock \emph{\bibinfo{journal}{npj Quantum Information}} \textbf{\bibinfo{volume}{7}}, \bibinfo{pages}{73} (2021).

\bibitem{Agresti2024}
\bibinfo{author}{Agresti, I.} \textit{et al.}
\bibinfo{title}{Demonstration of hardware efficient photonic variational quantum algorithm.}
\newblock \emph{\bibinfo{journal}{arXiv preprint}} arXiv:2408.10339 (2024).

\bibitem{chandarana2023digitized}
\bibinfo{author}{Chandarana, P.} \textit{et al.}
\bibinfo{title}{Digitized counterdiabatic quantum algorithm for protein folding.}
\newblock \emph{\bibinfo{journal}{Physical Review Applied}} \textbf{\bibinfo{volume}{20}}, \bibinfo{pages}{014024} (2023).

\bibitem{clements2016optimal}
\bibinfo{author}{Clements, W. R.} \textit{et al.}
\bibinfo{title}{Optimal design for universal multiport interferometers.}
\newblock \emph{\bibinfo{journal}{Optica}} \textbf{\bibinfo{volume}{3}}, \bibinfo{pages}{1460--1465} (2016).









\end{thebibliography}

\end{document}